# Orientational ordering and close packing properties of quasi-one-dimensional hard Gaussian overlap particles

*Sakineh Mizani[a)], Péter Gurin[b)] and Szabolcs Varga[b)]*

[a)]*Institute for Applied Physics, University of Tübingen, Auf der Morgenstelle 10, 72076 Tübingen, Germany*

[b)]*Physics Department, Centre for Natural Sciences, University of Pannonia, PO Box 158, Veszprém, H-8201 Hungary*

## Abstract

We investigate the orientational ordering and close-packing behavior of hard Gaussian overlap (HGO) particles, which are confined into a quasi-one-dimensional (q1D) channel. In the channel, particles are allowed to move along the channel and to rotate in three dimensions. Using the transfer operator method, we show that oblate particles align with their short axes along the channel, while prolate particles favor planar alignment perpendicular to the axis of the channel. While perfect orientational ordering develops in the fluid of oblate particles, the ordering is just partial in the fluid of prolate ones even at the close packing density. The pressure ratio of freely rotating and parallel particles ($P/P_\parallel$), which is an effective marker of structural changes, exhibits a single peak for oblate particles and no peak for prolate ones with increasing density. The close-packing behavior is characterized by exponents for the divergence of pressure ($P \sim \alpha \, P_{//}$), the decay of orientational fluctuations ($<(\theta_p - \theta)^2> \sim P^\beta$), and the behavior of the orientational correlation length ($\xi \sim P^\gamma$). The obtained values are $\beta = -1$, $\gamma = 0$ for both oblate and prolate particles, while $\alpha = 2$ for oblate and $\alpha = 1.5$ for prolate particles. Moreover, prolate particles belong to the universality class of hard superellipses, where the combinations $\alpha + \beta = 1/2$ and $\beta + \gamma = -1$ hold exactly for any $k > 1$ (S. Mizani et al., Phys. Rev. E 111, 064121 (2025)). However, oblate particles do not belong to this universal class because $\alpha + \beta = 1$.

## Introduction

The self-assembly of colloidal particles into ordered structures is a cornerstone of materials science, enabling applications from photonic crystals to drug delivery systems [1-5]. Important factors in this process are the particle shape and the spatial confinement, as both profoundly influence both positional (translational) and orientational (rotational) ordering. Non-spherical shapes introduce anisotropy, favoring mesophases such as nematic and smectic phases over isotropic fluids, as excluded volume interactions are strongly directional [6-12]. Confinement, by contrast, restricts accessible volume, suppressing entropy and profoundly altering phase and structural behavior. Depending on geometry and density, it can stabilize layered or crystalline structures, but may also induce frustration, re-entrant melting, or glassy dynamics [13-16]. Together, shape and confinement create a rich phase behaviour, where even subtle changes in geometry or dimensionality can yield unexpected structures. Understanding these effects is crucial for designing responsive nanomaterials, yet their interplay remains underexplored in reduced dimensions [17, 18].

In three-dimensional (3D) bulk systems, the ordering of spherical particles—modeled as hard spheres—has been extensively investigated, providing a well-established baseline for understanding the effects of excluded volume interactions on the positional ordering of colloidal suspensions [19-21]. Spherical hard bodies in bulk exhibit a classic fluid-to-crystal transition at a packing fraction of approximately 0.494, forming face-centered cubic (fcc) lattice driven purely by entropy [22-24]. Introducing particle non-sphericity disrupts this simple melting process. For example, several mesophases such as the nematic, smectic, and columnar phases can be stabilized with increasing density if the shape of the particle is sufficiently anisotropic [25-27]. Moreover, increasing shape anisotropy enhances orientational correlations among particles, leading to the destabilization of isotropic phase with respect to nematic phase in agreement with theoretical predictions based on excluded-volume considerations [28]. However, the stability of more ordered mesophases, such as smectic and columnar phases, is less sensitive to changes in shape anisotropy [29]. Nonetheless, systems of nearly spherical particles often remain disordered over wide density ranges, showing only limited local ordering even at high packing fractions [30]. These bulk behaviors collectively demonstrate that shape anisotropy significantly shifts the balance between translational and rotational degrees of freedom, highlighting the intricate interplay of orientational and packing entropies that governs the ordering properties of colloidal particles [31-34].

Reducing spatial dimensionality to two dimensions (2D) while preserving full three-dimensional (3D) rotational freedom introduces novel phase behavior, particularly for colloids adsorbed at liquid-liquid interfaces [35-40]. Early models of hard spherocylinders with midpoints confined to a plane showed that in-plane positional confinement still allows significant orientational ordering [41]. Spherical colloidal particles at a flat fluid interface assemble into two-dimensional hexagonal crystals, whose mechanical response depends on both the area fraction of the particles and the direction of applied forces [42]. Non-spherical particles amplify orientation effects: prolate particles stand up (tilt out of the plane), while oblate particles tilt into the plane with their axis of symmetry to minimize the area they occupy in the plane [43-45]. As a result, in-plane isotropic, nematic and 2D crystalline ordering arises with increasing densities [46, 47]. In addition to this, orientational-positional mismatch can lead to orientation-dependent interactions at interfaces [48]. In related 2D colloidal systems, this interplay can give rise to intermediate hexatic phases that display bond-orientational order while translational order remains partially disrupted [49-50]. These interfacial systems emphasize how interfacial forces (such as contact-line pinning) couple translational and rotational motion of particles at fluid interfaces [51, 52].

Even stronger confinement can be achieved by placing colloidal particles into very narrow nanochannels. Such systems are often referred to as quasi-one-dimensional (q1D) because the particles form a single file, where they cannot pass each other in the row [53, 54]. In such environments, the static and dynamic properties of spherical particles have some special features, as long-range positional order cannot develop. One of the special features is the lack of a true phase transition as the interaction is short-range between colloidal particles [55]. Apart from this fact, structural changes can emerge upon compression as the close packing structure of the system can be very complex [56, 57]. For example, spherical colloids can adopt zigzag, helical, or even layered close packing structures in cylindrical and rectangular channels [58]. In the lack of out-of-line positional freedom, the system becomes strictly one-dimensional (1D) and positional ordering can evolve along the channel only. If the interaction is short-range, the structure and the thermodynamics of the system can be determined analytically [59-61]. In such 1D systems, structural change is not feasible as the particles form a row at any density. However, the short-range and long-range behavior of the positional arrangement can be different in 1D channels. A recent study of Bouzar and Messina [62] has shown that hard sphere colloids form such a solid structure in 1D confinement, where the decay of positional correlations is algebraic inside and exponential outside the solid domain. Moreover, the size of the algebraic solid domain diverges at the close packing density.

The phase behavior of 1D hard body fluids becomes more complex with the deformation of the particle's shape from spherical to non-spherical, because the arising orientation-dependent anisotropic interactions give rise to orientational ordering. These systems are q1D, because one positional freedom is accompanied by one or two orientational ones. The simplest q1D systems are two dimensional objects such as the ellipse and the rectangle, which are confined to a straight line and are allowed to rotate freely in 2D. These systems exhibit orientational structural change and peculiar close packing properties [63-67]. The structural change may arise between quasi-isotropic and orientationally ordered structures, which can be tetratic, nematic, or even mixed [66, 68]. The close packing property can be characterized with three exponents: $\alpha$ measures the contribution of angular and positional fluctuations to the pressure, $\beta$ characterizes the decay of mean square orientational fluctuations, and $\gamma$ describes the divergence of the orientational correlation length with the pressure. Studies of these quantities in q1D superdisk and superellipse systems found that they exhibit deformation dependence but no shape anisotropy dependence [66, 67]. Moreover, some combinations of these quantities are universal in the sense that $\alpha+\beta=1/2$ and $\beta+\gamma=-1$ for 2D objects with one orientational freedom [66, 67]. However, the effect of orientational freedoms on the orientational ordering properties is much less known for 3D particles confined to a straight line. In these systems, the particles are allowed to perform 3D rotations, which means that two orientational freedoms (fluctuations in both polar and azimuthal angles) are present. In such systems, we have only limited knowledge about the close packing behavior of the pressure for some models [63, 68-69]. For example, the contribution of the additional orientational freedom, which is the polar angle, is 0.5 (1) to $\alpha$ for ellipsoidal (rectangular) type shapes [63]. However, our knowledge is still very limited about the possible structural changes at intermediate densities and the close packing behavior of orientational fluctuations and correlations. Furthermore, it is still unclear how the shape of particles affects their ordering properties, as 3D particles can be either prolate or oblate.

In this work, we investigate the impact of the 3D rotation and the particle's shape on the ordering of hard colloidal particles confined into nanochannels. Using the transfer operator method, we determine the orientational ordering properties of both prolate and oblate particles. We examine how a particle's aspect ratio governs the stability of quasi-isotropic and nematic structures and the close packing behavior. To do this, we use the hard Gaussian overlap model [70-73] for describing the interaction between the colloidal particles. The advantage of this model is that it is simple and incorporates both the oblate and prolate shapes. Moreover, we will show that even some analytical results can be gained with this model in the vicinity of close packing density.

Our results point out that while prolate particles adopt a perpendicular alignment to the channel axis, oblate particles align along the channel. The different ordering strategy manifests in partial ordering of prolate particles and perfect orientational ordering of oblate ones with approaching the close packing density. Interestingly, the close packing properties of both shapes do not depend on the aspect ratio of the particle, even if their close packing behavior is different. We will show that $\alpha$ is discontinuous at the spherical shape as $\alpha=2$ for oblate particles, while $\alpha=1.5$ for prolate ones.

## Model

We consider a quasi-one-dimensional (q1D) model of $N$ particles whose centers are constrained to move along a straight line of length $L$ along the $z$-axis, while each particle can rotate freely in three dimensions. The particle interactions are described by the hard Gaussian overlap (HGO) model [70-72], originally introduced by Berne and Pechukas [73], where overlaps are strictly forbidden between particles. Here, we must note that HGO particles do not have well-defined volume and surface area, as only the contact distance between two HGO particles is defined. Apart from this, HGO contact distance represents quite accurately the contact distance between two uniaxial hard ellipsoids [74]. Therefore HGO model is often considered as an analytical approximation of hard ellipsoid particles [75]. Figure 1 illustrates the case when two HGO particles are in contact. Using the orientational unit vectors ($\vec{\Omega}_1$ and $\vec{\Omega}_2$) and centre-to-centre unit vector ($\vec{\Omega}_{12}$) between two neighboring particles, the contact distance, which corresponds to the possible shortest distance between two HGO particles, is defined by

$$\sigma\left(\vec{\Omega}_1,\vec{\Omega}_2,\vec{\Omega}_{12}\right)=\frac{\sigma_{ss}}{\sqrt{1-\frac{\chi}{2}\left\{\frac{\left(\vec{\Omega}_1\cdot\vec{\Omega}_{12}+\vec{\Omega}_2\cdot\vec{\Omega}_{12}\right)^2}{1+\chi\vec{\Omega}_1\cdot\vec{\Omega}_2}+\frac{\left(\vec{\Omega}_1\cdot\vec{\Omega}_{12}-\vec{\Omega}_2\cdot\vec{\Omega}_{12}\right)^2}{1-\chi\vec{\Omega}_1\cdot\vec{\Omega}_2}\right\}}},\tag{1}$$

where $\chi=(k^2-1)/(k^2+1)$ and $k=\sigma_{ee}/\sigma_{ss}$ is the particle's aspect ratio. Here, $\sigma_{ss}$ is the side-by-side contact distance (equatorial diameter) and $\sigma_{ee}$ denotes the end-to-end contact distance (polar diameter). Note that $k<1$ corresponds to an oblate (pancake) shape, while $k>1$ to a prolate (cigar) shape, and $k=1$ to a sphere. In a spherical coordinate frame, the orientational unit vector of a particle $i$ is given by $\vec{\Omega}_i=\left(\sin\theta_i\cos\varphi_i,\sin\theta_i\sin\varphi_i,\cos\theta_i\right)$, where the polar ($\theta_i$) and the azimuthal ($\varphi_i$) angles are in the intervals of $0\leq\theta_i\leq\pi$ and $0\leq\varphi_i\leq2\pi$. Moreover, $\vec{\Omega}_{12}=\left(0,0,1\right)$ is due to the one-dimensional positional confinement. Substituting $\vec{\Omega}_1$, $\vec{\Omega}_2$ and $\vec{\Omega}_{12}$ into Equation (1), the HGO contact distance for adjacent particles simplifies to

$$\sigma\left(\theta_1,\theta_2,\varphi_{12}\right)=\frac{\sigma_{ss}}{\sqrt{1-\frac{\chi}{1-\chi^2\cos^2\gamma}\left\{\cos^2\theta_1+\cos^2\theta_2-2\chi\cos\gamma\cos\theta_1\cos\theta_2\right\}}},\tag{2}$$

where $\cos\gamma = \vec{\Omega}_1\vec{\Omega}_2 = \sin\theta_1\sin\theta_2\cos(\varphi_{12}) + \cos\theta_1\cos\theta_2$ and $\varphi_{12} = \varphi_1 - \varphi_2$. This orientation-dependent contact distance serves as the input for the transfer operator method, which will be presented in the Theory section. Note that $\sigma$ depends on $\theta_1$, $\theta_2$ and $\varphi_{12} = \varphi_1 - \varphi_2$.

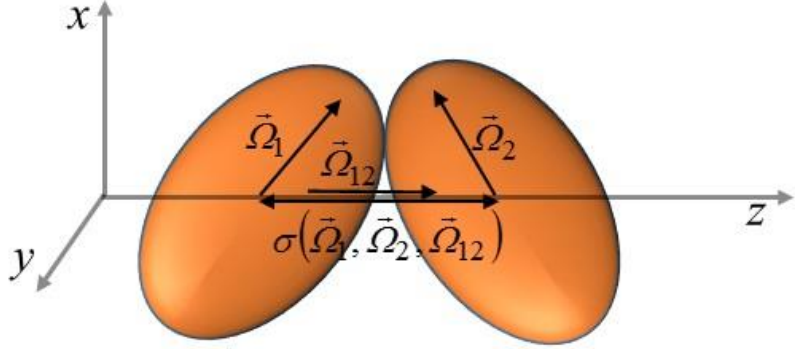

**Figure 1.** Schematic representation of the hard Gaussian overlap (HGO) model, where the centers of the particles are confined to the $z$-axis. Two prolate particles are in contact, with their orientations represented by unit vectors $\vec{\Omega}_1$ and $\vec{\Omega}_2$. The center-to-center unit vector connecting the particles is denoted $\vec{\Omega}_{12}$. The contact distance ($\sigma$) corresponds to the distance between the centers of two particles.

Since overlaps between HGO particles are forbidden, the contact distance governs the orientational ordering of the particles. At close packing, all particles are in contact, and the contact distance between neighboring particles must take its minimum value. For oblate particles ($k<1$), this minimum occurs when all particles align along the $z$-axis ($\theta_i=0$). Substituting $\theta_1=0$ and $\theta_2=0$ into Equation (2), we get that

$$\sigma\left(\theta_1 = 0, \theta_2 = 0, \varphi_{12}\right) = \frac{\sigma_{ss}}{\sqrt{\dfrac{1-\chi}{1+\chi}}} = \sigma_{ee}, \tag{3}$$

which is really the possible shortest distance between two oblate particles. Consequently, oblate particles tend to form a nematic phase with their orientational unit vectors parallel to the $z$-axis. In contrast, for prolate particles ($k>1$), we get the minimum contact distance at $\theta_1 = \theta_2 = \pi/2$ as

$$\sigma\left(\theta_1 = \frac{\pi}{2}, \theta_2 = \frac{\pi}{2}, \varphi_{12}\right) = \sigma_{ss}, \tag{4}$$

which corresponds to the side-by-side arrangement between two HGO particles. As a result, prolate particles tend to adopt planar alignment with respect to the $z$-axis, forming a planar nematic phase rather than a conventional nematic phase.

In the vicinity of close-packing density, where particles align either along the channel or perpendicular to it, the contact distance can be approximated using the Taylor expansion. This expansion is performed up to the second order in $\theta_1$ and $\theta_2$ around $\theta_1 = 0$ and $\theta_2 = 0$ for oblate particles, which can be written as

$$\sigma\left(\theta_1, \theta_2, \varphi_{12}\right) \approx \sigma_{ee}\left(1 - \frac{\chi}{2}\frac{\theta_1^2 + \theta_2^2}{1 - \chi^2} - \frac{\cos\varphi_{12}\chi^2\theta_1\theta_2}{1 - \chi^2}\right). \tag{5}$$

The same second order procedure around $\theta_1 = \frac{\pi}{2}$ and $\theta_2 = \frac{\pi}{2}$ provides the following approximate contact distance for prolate particles:

$$\sigma\left(\theta_1, \theta_2, \varphi_{12}\right) \approx \sigma_{ss}\left(1 + \frac{\chi}{2}\frac{\left(\pi/2 - \theta_1\right)^2 + \left(\pi/2 - \theta_2\right)^2}{1 - \chi^2\cos^2\varphi_{12}} - \frac{\cos\varphi_{12}\chi^2\left(\pi/2 - \theta_1\right)\left(\pi/2 - \theta_2\right)}{1 - \chi^2\cos^2\varphi_{12}}\right). \tag{6}$$

We will use these second-order contact distances for the analytical analysis of the closing packing properties of q1D HGO fluid.

**Theory**

To determine the thermodynamic and structural properties of q1D HGO particles, we employ the transfer operator method (TOM). This formalism is devised in an isobaric (*NPT*) ensemble for systems where the pair interactions are short ranged and the positional freedoms are strongly restricted [76]. The method is particularly effective for systems characterized by first-neighbor hard body interactions, where the statistical mechanical problem can be reformulated with the help of the contact distance between adjacent particles. In the q1D channel, the key quantity of HGO particles is the following function

$$K_0\left(\theta_1, \theta_2, \varphi_{12}\right) = \frac{\exp[-P\sigma\left(\theta_1, \theta_2, \varphi_{12}\right)]}{P} \tag{7}$$

where $P = p/k_B T$, $p$ is the axial pressure, $k_B$ is the Boltzmann-factor, and $T$ is the temperature. Using this kernel function, we can define the following eigenvalue equation,

$$\int d\varphi_2 d\theta_2 \sin\theta_2 K_0(\theta_1, \theta_2, \varphi_{12})\psi\left(\theta_2, \varphi_2\right) = \lambda\psi\left(\theta_1, \varphi_1\right) \tag{8a}$$

where $\lambda$ is the eigenvalue and $\psi$ is the corresponding eigenfunction. The integrations are performed over the full solid angle, i.e., $0 \leq \theta < \pi$ and $0 \leq \varphi < 2\pi$. In addition, this eigenvalue equation is supplemented with the following normalization condition

$$\int d\varphi d\theta \sin\theta \psi^2(\theta,\varphi) = 1 \qquad (8b)$$

The solution of the eigenvalue-problem (Equations (8a) and (8b)) yields a set of eigenvalues $\lambda_0 > |\lambda_1| > |\lambda_2| > \ldots$ and corresponding eigenfunctions $\psi_i(\theta,\varphi)$, where the largest eigenvalue ($\lambda_0$) can be used to determine the equilibrium thermodynamic quantities, while the associated eigenfunction provides the orientational distribution function (ODF) as follows $f(\theta,\varphi) = \psi_0^2(\theta,\varphi)$. As the HGO contact distance (Equation (2)) is non-additive, it is not feasible to solve the eigenvalue problem analytically. Therefore, we solve Equations (8a) and (8b) with the discretization of $\theta$ and $\varphi$ angles as follows: $\psi = \psi(\theta_i,\varphi_i)$, where $\theta_i = i \Delta\theta$ and $\varphi_i = j \Delta\varphi$ with $\Delta\theta = \pi/200$ and $\Delta\varphi = \pi/100$. The integrations are performed with the trapezoidal quadrature, and the successive substitution method is used to get the solution of the eigenvalue-problem. As an initial guess for the eigenfunction, we use the following constant and normalized function, which is given by $\psi = 1/\sqrt{4\pi}$. Using this initial guess, we always get the largest eigenvalue with the numerical method. However, due to the azimuthal angle invariance of the HGO contact distance (see Equation (2)) as it depends on $\varphi_{12}$, the eigenfunction belonging to the largest eigenvalue is independent of $\varphi$, i.e., $\psi_0 = \psi_0(\theta)$. To get $\lambda_1$ and the corresponding eigenfunction, we project out the largest eigenvalue solution. To do this, we only have to replace the kernel function (Equation 7) with the following new kernel:

$$K_1(\theta_1,\theta_2,\varphi_{12}) = K_0(\theta_1,\theta_2,\varphi_{12}) - \lambda_0 \psi_0(\theta_1)\psi_0(\theta_2) \qquad (9)$$

The numerical solution of Equations (8a) and (8b) with this new kernel provides the second largest eigenvalue ($\lambda_1$) together with $\psi_1$ depending on both $\theta$ and $\varphi$. We use these results to get some information about the orientational correlations. For example, the orientational correlation length can be obtained from $\xi = 1/\ln(\lambda_0/|\lambda_1|)$ [64]. As the Gibbs free energy ($G$) is related to $\lambda_0$ as follows: $G = - Nk_BT \ln(\lambda_0)$, the equation of state can be obtained from $1/\rho = \partial(G/N)/\partial p$, where $\rho = N/L$ is the 1D number density. Using the ODF, which has only $\theta$ dependence for HGO particles, i.e., $f(\theta) = \psi_0^2(\theta)$, the nematic order parameter is the orientational average of the second Legendre-polynomial ($P_2 = 1.5\cos^2\theta - 0.5$), i.e.,

$$S = \int d\varphi \int d\theta \sin\theta f(\theta) P_2(\cos(\theta)) \qquad (10)$$

It is also useful to determine the average of orientational fluctuations from the preferred orientation. We will show in the Result section that the preferred orientations are $\theta_p = 0$ for oblate shapes and $\theta_p = \pi/2$ for prolate ones. Therefore, the average of the mean square orientational fluctuations is given by

$$<(\theta_p - \theta)^2> = \int d\varphi \int d\theta \sin\theta f(\theta)(\theta_p - \theta)^2 \qquad (11)$$

To get some insight into the phase behavior of q1D HGO model, we examine such additive hard body models, which can be described by the following contact distance

$$\sigma\left(\theta_1,\varphi_1,\theta_2,\varphi_2\right) = \frac{\sigma(\theta_1,\varphi_1)+\sigma(\theta_2,\varphi_2)}{2} \tag{12}$$

Substituting Equation (12) into Equation (7) and then into Equation (8a) results in

$$\frac{e^{-P\sigma(\theta_1,\varphi_1)/2}}{P}\int d\varphi_2 d\theta_2 \sin\theta_2 e^{-P\sigma(\theta_2,\varphi_2)/2}\psi\left(\theta_2,\varphi_2\right) = \lambda\psi\left(\theta_1,\varphi_1\right) \tag{13}$$

As the integral in Equation (13) is an orientation independent constant, the orientation dependent prefactor prescribes the orientation dependence of the eigenfunction. Consequently, $\psi\left(\theta,\varphi\right) = ce^{-P\sigma(\theta,\varphi)/2}$, where $c$ is a normalization prefactor. Substituting this function into Equation (13), we get that

$$\lambda = \int d\varphi d\theta \sin\theta \frac{e^{-P\sigma(\theta,\varphi)}}{P} \tag{14}$$

This is actually the largest eigenvalue of Equation (13), i.e., $\lambda_0=\lambda$. Applying the normalization condition (Equation (8b)), we can determine $c$, and we can write that

$$\psi\left(\theta,\varphi\right) = \frac{e^{-P\sigma(\theta,\varphi)/2}}{\sqrt{\int d\varphi_1 \int d\theta_1 \sin\theta_1 e^{-P\sigma(\theta,\varphi_1)}}} \tag{15}$$

and

$$f\left(\theta,\varphi\right) = \frac{e^{-P\sigma(\theta,\varphi)}}{\int d\varphi_1 \int d\theta_1 \sin\theta_1 e^{-P\sigma(\theta,\varphi_1)}} \tag{16}$$

This equation informs us that additive systems order along that special orientation, which gives the maximum of Equation (16). This occurs at the minimum of $\sigma(\theta,\varphi)$. Moreover, using $G=-Nk_BT\ln(\lambda_0)$ and $1/\rho=\partial(G/N)/\partial p$, the following relationship can be obtained between $\rho$ and $P$

$$\frac{1}{\rho} = \frac{1}{P} + \int d\varphi d\theta \sin\theta f\left(\theta,\varphi\right)\sigma\left(\theta,\varphi\right) \tag{17}$$

From this result, we can see that only the minimal value of $\sigma(\theta,\varphi)$ contributes at very large pressures, i.e., $1/\rho=\min(\sigma(\theta,\varphi))$. Therefore the close packing density of the additive systems is given by $\rho_{cp}=1/\min(\sigma(\theta,\varphi))$. Moreover, if the orientations of particles are frozen into that special orientation, which belongs to the minimum of $\sigma(\theta,\varphi)$, i.e., the system consists of parallel particles,

Equation (17) simplifies to $1/\rho = 1/P + \min(\sigma(\theta,\varphi))$. Renaming the pressure of orientationally frozen particles as $P_{\parallel}$ we can also write that

$$P_{\parallel} = \rho/(1 - \rho \min(\sigma(\theta,\varphi))) \qquad (18)$$

Note that this equation is identical to the equation of state of 1D hard rods if the contact distance is constant for all orientations [77]. For comparison, the isotropic reference pressure is defined as $P_{iso} = \rho/(1 - \rho \langle \sigma \rangle)$, where $\langle \sigma \rangle = \int d\theta_1 \sin\theta_1 \int d\theta_2 \sin\theta_2 \int d\varphi_{12} \sigma(\theta_1, \theta_2, \varphi_{12})/8\pi$. Furthermore, it is also an interesting result that $\lambda_1 = 0$, because it follows from Equation (9) that $K_1 = 0$ for additive contact distances. Consequence of this result is that, the additive hard systems do not exhibit long-range orientational correlations, because it follows from $\xi = 1/\ln(\lambda_0/\lambda_1)$, that $\xi$ is always zero. Therefore, additive systems are orientationally uncorrelated in the entire density range.

Finally, we mention that the pressure and the density are made dimensionless as follows: $P^* = P\sigma_{ss}$ and $\rho^* = \rho d$, where $d = \min(\sigma_{ee}, \sigma_{ss})$. With this choice, the range of the dimensionless density is set to be the same for both prolate and oblate shapes as $0 < \rho^* < 1$.

## Results

In this section, we present the orientational ordering and close-packing properties of quasi-one-dimensional (q1D) hard Gaussian overlap (HGO) fluids coming from the exact TOM. The numerical findings are interpreted in the light of analytically tractable additive hard-body models, which provide physical insight into the nature of HGO interactions. The comparison indicates that the contact distance between two HGO particles becomes effectively additive near the close-packing limit, as orientational fluctuations around the preferred alignment become very small. The TOM study further enables a direct comparison of results coming from oblate and prolate shapes, revealing markedly different orientational responses under confinement.

Figure 2 illustrates the evolution of ODF ($f(\theta)$) with increasing pressure for two representative shape anisotropies: oblate HGO particles ($k=1/3$, solid lines) and prolate HGO particles ($k=3$, dashed lines). At the lowest pressure ($P^*=1$) both curves are nearly flat, reflecting an almost isotropic orientational distribution. As the pressure rises ($P^*=5$ then 10), clear maxima and orientational preferences emerge. The oblate HGO particles preferentially align around $\theta_p=0$, whereas the prolate ones exhibit maxima near $\theta_p=\pi/2$. The increasing peak amplitudes with $P^*$ demonstrate a progressive enhancement of orientational order towards $\theta_p$ as density (or pressure) increases. The origin of this distinct orientational behavior for oblate and prolate HGO particles can be rationalized through a simple packing argument. Both types of particles tend to minimize their occupied lengths along the $z$-axis, yet they achieve this through opposite orientational strategies. Oblate HGO particles reduce their lengths along the $z$-axis most effectively when their symmetry axes align along the $z$-axis, which is in agreement with Equation (3). In contrast, prolate HGO particles minimize their $z$-projection lengths, when their long symmetry axes are in the $xy$-plane in accordance with Equation (4). Consequently, close packing is accomplished by parallel stacking of oblate particles along the $z$-axis, but through isotropic in-plane arrangement of prolate ones in the $xy$-plane. With increasing pressure, these tendencies become more pronounced, resulting in the progressive sharpening of the peak in ODF (see Figure 2). To gain analytical insight into the high-pressure (close packing) regime, we adopt the following additive approximation for the contact distance between neighboring oblate HGO particles,

$$\sigma\left(\theta_1,\theta_2\right) \approx \sigma_{ee}\left(1 - \frac{\chi(\theta_1^2 + \theta_2^2)}{2}\right) \tag{19}$$

which can be obtained from Equation (5), by considering the limit of very weak non-sphericity, in which only the first-order term in the shape parameter $\chi$ is retained. In this expansion, the cross term proportional to $\theta_1\theta_2$ is missing, yielding a contact distance that depends additively on the orientations of individual particles. This additive form allows the contact distance to be expressed as a sum of single-particle contributions, thereby decoupling the orientational dependence of neighboring particles. For the sake of brevity, we use $a = \sigma_{ee}$ and $b = -\chi \sigma_{ee}/2$, which leads to a simple reduced form

$$\sigma\left(\theta_1, \theta_2\right) = a + b\left(\theta_1^2 + \theta_2^2\right), \tag{20}$$

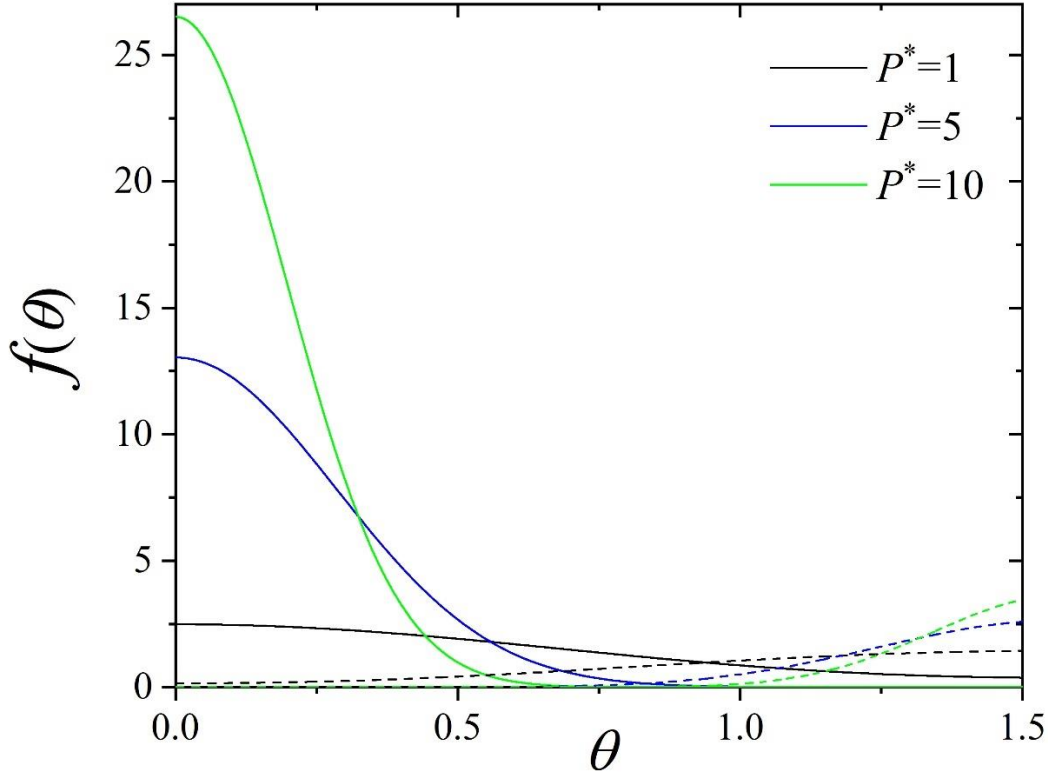

**Figure 2.** Effect of increasing pressure ($P^*=P\sigma_{ss}$) on orientational distribution function for prolate ($k=3$) and oblate ($k=1/3$) HGO particles. Solid lines show results for $k=1/3$, while dashed lines for $k=3$.

Note that this contact distance does not depend on the azimuthal angle ($\varphi_{12}$). With the comparison of Equations (12) and (20) we get that $\sigma(\theta)=a+2b\theta^2$. Substituting this parameter into Equation (16) and assuming very high pressures, we get the following analytical results for the ODF

$$f\left(\theta\right) = \frac{e^{-2Pb\theta^2}}{4\pi \int\limits_0^{\pi/2} d\theta \sin\theta\, e^{-2Pb\theta^2}} \approx Pb\frac{e^{-2Pb\theta^2}}{\pi} \tag{21}$$

which reproduces the numerical trends, specifically the gradual sharpening of the ODF with increasing pressure. A similar approach applies to prolate particles, for which the relevant close-packing fluctuations occur near $\theta_p=\pi/2$. Considering only very weakly anisotropic shapes ($\chi\approx0$) in Equation (6), the first order expansion in $\chi$ for the prolate contact distance becomes

$$\sigma\left(\theta_1, \theta_2\right) \approx \sigma_{ss}\left(1 + \frac{\chi}{2}\sum_{i=1}^{2}\left(\pi/2 - \theta_i\right)^2\right). \tag{22}$$

Note that this prolate contact distance is additive and independent of the azimuthal angle ($\varphi_{12}$). Moreover, it can be rewritten as $\sigma(\theta_1,\theta_2)=a+b$ $((\pi/2-\theta_1)^2+((\pi/2-\theta_2)^2)$, where $a=\sigma_{ss}$ and $b=\sigma_{ss}\chi/2$. Inserting $\sigma(\theta)=a+2b$ $((\pi/2-\theta)^2$ into Equation (16), we get that the ODF is given by

$$f\left(\theta\right) = \frac{e^{-2Pb(\pi/2-\theta)^2}}{4\pi \int\limits_0^{\pi/2} d\theta \sin\theta\, e^{-2Pb(\pi/2-\theta)^2}} = \frac{\sqrt{Pb}}{\sqrt{2}\pi^{1.5}}e^{-2Pb\left(\frac{\pi}{2}-\theta\right)} \tag{23}$$

which has maximum at $\theta=\pi/2$. Interestingly, the pressure dependence of ODF is different for oblate and prolate shapes, because $f(\theta=\pi/2)\sim P^{1/2}$ for $k>1$, while $f(\theta=0)\sim P$ for $k<1$. These analytical results agree well with the numerical results shown in Figure 2, where it can be seen clearly that the pressure dependence of the ODF differs between oblate and prolate shapes. Therefore, the additive approximation for the contact distance explains the sharper peak for oblate particles (stronger alignment along the $z$-axis) compared to the broader, more slowly growing peak for prolate particles (gradual tilting into the $xy$-plane), as seen in Figure 2.

To see the extent of orientational ordering, the nematic order parameter (see Equation (10)) is shown as a function of density in Figure 3. Oblate HGO particles show a steady increase of $S$ as a function of density approaching the close packing density ($\rho^*=1$). This shows that a perfectly ordered nematic structure evolves at the close packing. In contrast, prolate particles display decreasing $S$ values tending towards $-1/2$, corresponding to an isotropic in-plane distribution. This means that there is no in-plane orientational order even at the close packing density in the perpendicular plane given by $\theta=\pi/2$. Such a structure is often referred to as a planar arrangement, because the particles align isotropically into a plane instead of aligning along a nematic director. It can be shown that the oblate and prolate close packing limits of the order parameter can be obtained analytically by inserting Equations (21) and (23) into Equation (10). Taking the high pressure limit, $S$ becomes 1 for oblate particles, and $S$ goes to $-1/2$ for prolate ones. As $S$ changes from convex to concave shape with increasing density, oblate particles undergo an isotropic–nematic structural change at intermediate densities and become perfectly ordered at the close packing density. Contrary to this, $S$ is always concave for prolate particles, which means that the planar ordering evolves very slowly with increasing density without reaching perfect orientationally order at the close packing density.

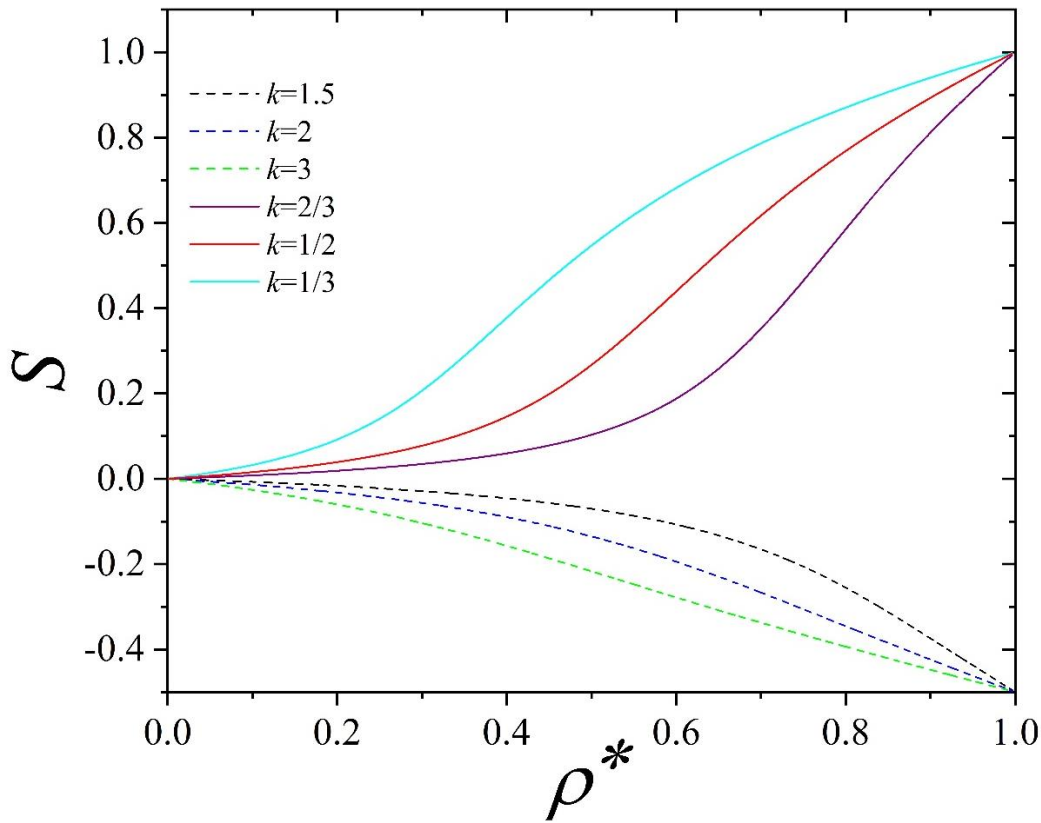

**Figure 3.** Nematic order parameter ($S$) as a function of density ($\rho*=\rho d$) for various aspect ratios ($k$).

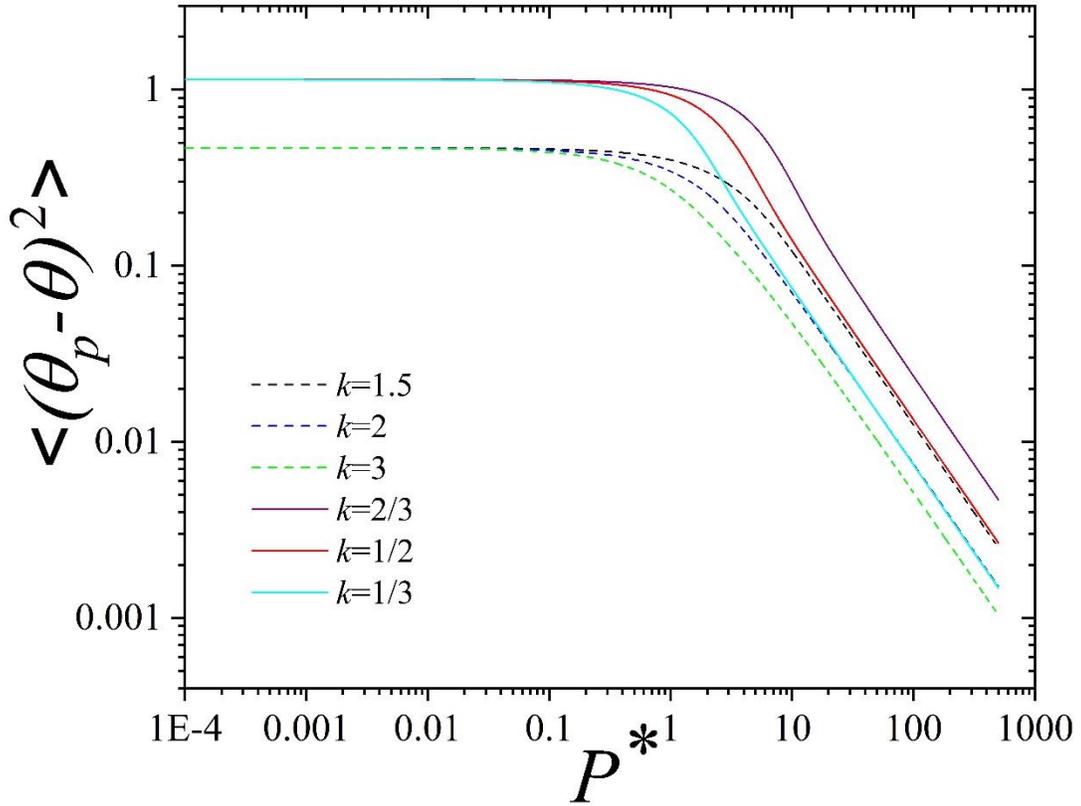

**Figure 4.** Average angular fluctuation as a function of pressure ($P^*=P\sigma_{ss}$) for various aspect ratios ($k$).

The pressure dependence of the average of orientational fluctuation ($<(\theta_p-\theta)^2>$) is shown in Figure 4. Two distinct regimes are observed: at low pressure, fluctuations remain nearly constant, reflecting weak orientational order as the structure is quasi-isotropic. At intermediate pressures, the fluctuations decrease significantly as orientational order develops. At high pressures, the average of orientational fluctuation follows a power-law decay, $<(\theta_p-\theta)^2>\sim P^\beta$, defining the exponent $\beta$. Our numerical calculations show that $\beta$=-1 for both oblate and prolate shapes. The observed high-pressure power-law decay of orientational fluctuations is well captured within the additive approximation. For oblate HGO particles, where $\theta_p$=0, Equations (11) and (21) lead to the expression for average angular fluctuations,

$$\left\langle (\theta_p-\theta)^2 \right\rangle \approx \frac{1}{2}\frac{1}{Pb} \tag{24}$$

Similarly, for prolate HGO particles, where $\theta_p=\pi/2$, substituting the analytical ODF (Equation (23)) into Equation (11) yields the same mean-square fluctuation as given by Equation (24). From these results, we can see that the analytical additive models produce the same close-packing exponent for the orientational fluctuations, i.e., $\beta=-1$ for both prolate and oblate shapes. These results demonstrate that the effect of orientational fluctuations near close packing is universal and independent of the particle aspect ratio $k$. Despite the different preferred directions of oblate particles ($\theta_p=0$) and prolate ones ($\theta_p=\pi/2$), their orientational fluctuations decay with the same exponent at high pressures.

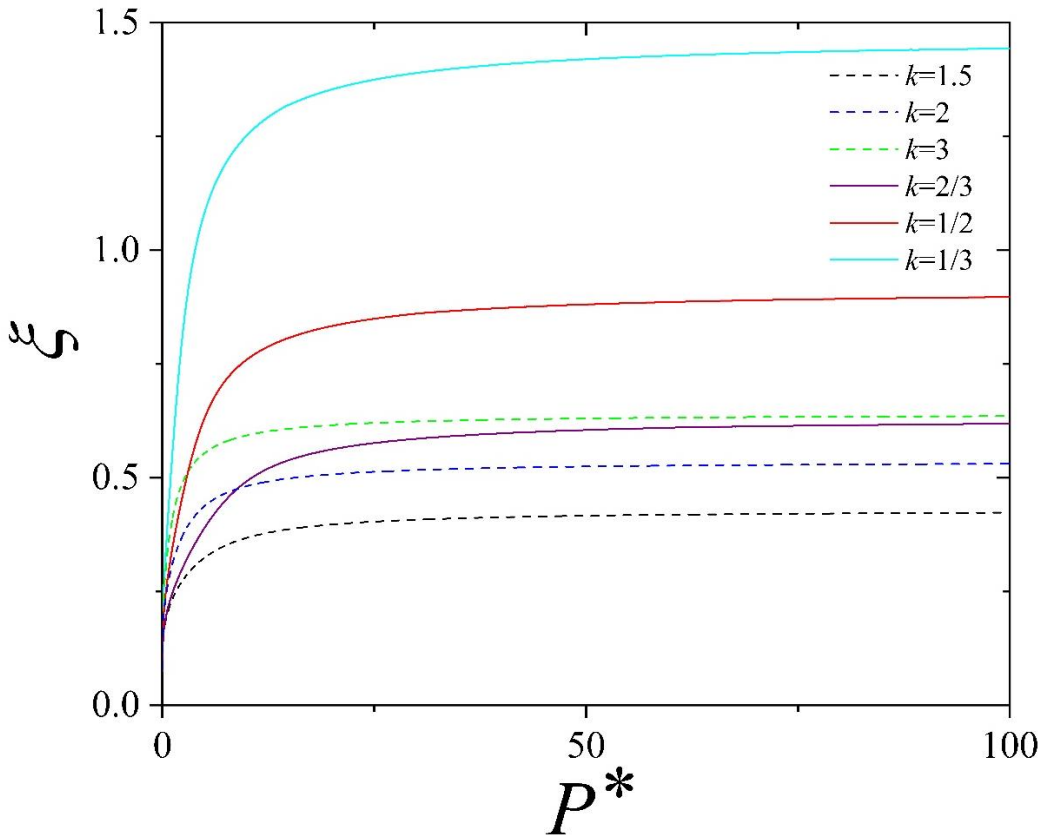

**Figure 5.** Orientational correlation length ($\xi$) as a function of pressure ($P^*=P\sigma_{ss}$) for various aspect ratios ($k$).

The orientational correlation length ($\xi$) informs us about the collective orientational motions of particles. If this quantity is short, the orientational fluctuation of a particle does not affect the orientational fluctuation of a far particle. This is the case for HGO particles, where long-range orientational correlations do not evolve even at the close packing density. We can see in Figure 5 that the correlation length is always very short and saturates at high pressures. Figure 5 also shows that the orientational correlation becomes stronger for more elongated ($k>1$) or more flat shapes ($k<1$) and the correlation length never goes beyond the second or third neighbor. These results indicate that the strong orientational ordering is due to the hard body exclusion between neighboring particles, while particles being far from each other do not have an impact on the orientational ordering. These results are in agreement with our findings for hard anisotropic particles interacting with an additive contact distance, where the orientational ordering can be partial or even very strong, and the correlation length, which is given by $\xi = 1/\ln(\lambda_0/\lambda_1)$, is zero as $\lambda_1 = 0$. We get the same result for HGO particles using the additive contact distance (see Equations (19) and (22)). The only difference between the numerical results presented in Figure 5 and the case of additive interaction is that $\xi$ increases first and saturates at high pressures, while $\xi$ of the additive system is zero at any pressure.

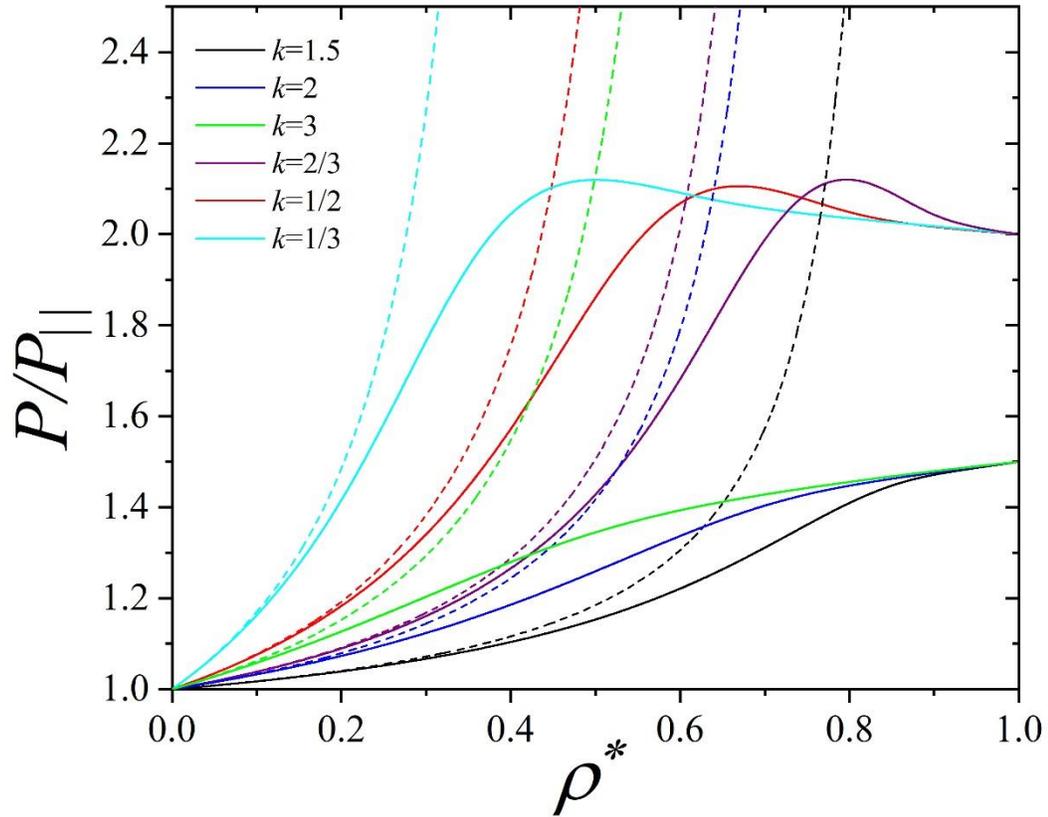

**Figure 6.** Pressure ratio ($P/P_{//}$) as a function of density ($\rho^* = \rho d$) for different aspect ratios ($k$). Solid lines represent exact results coming from TOM, while dashed lines represent the approximate isotropic pressure ratio ($P_{iso}/P_{//}$).

We get further information about the effect of orientational fluctuations by showing the density dependence of the pressure ratio $P/P_{//}$ for both prolate and oblate shapes. Figure 6 shows the exact numerical results for $P/P_{//}$ together with the isotropic pressure ratio ($P_{iso}/P_{//}$). In all cases, the pressure ratio remains larger than unity, indicating that orientational fluctuations enhance the pressure compared to the pressure of perfectly aligned particles. We can also see in Figure 6 that a pronounced difference emerges between oblate ($k<1$) and prolate ($k>1$) particles. For oblate particles, $P/P_{||}$ exhibits a maximum at intermediate density, marking the isotropic–nematic structural change as the convex-to-concave change of $S$. This behavior coincides with the divergence of the isotropic pressure, marking the stability limit of the isotropic structure. Moreover, the $P/P_{||}$ peak moves to the direction of lower densities with decreasing $k$, which indicates that the more anisotropic shape supports the formation of nematic structures with respect to the quasi-isotropic structure. In contrast, prolate particles show a smooth increase of $P/P_{||}$ without an intermediate peak, while the divergence in $P_{iso}/P_{||}$ is still present. The lack of an intermediate peak is the consequence of a weak tendency for orientational ordering, because even the close packing structure is only partially ordered with $S$=-1/2 and $S$ is always concave (see Figures (2) and (3)). Contrary to prolate particles, the tendency for quasi-isotropic to nematic structural change is much stronger in the fluid of oblate particles, where even perfect nematic order ($S$=1) evolves at the close packing density. The other interesting property of $P/P_{||}$ is its close packing value, which has a finite contribution from the orientational fluctuations. We can see in Figure 6 that $\alpha$=2 for $k$<1 and $\alpha$=1.5 for $k$>1, where $\alpha = \lim\limits_{\rho^* \to 1} P / P_{||}$. Note that, it is trivial to show for hard spheres ($k$=1 and $\sigma=\sigma_{ee}=\sigma_{ss}$) using Equations (17) and (18) that $\alpha$=1. Therefore, $\alpha$ is constant, but it becomes discontinuous at $k$=1. Moreover, we can conclude from these results that the orientational freedoms have different contributions to the pressure for oblate and prolate shapes. While the contributions of orientational and translational freedoms are the same to the pressure for oblate shapes, the contribution of orientational freedoms is just half of the contribution of translational freedom to the pressure for prolate shapes at the close packing density. It is also evident that only the translational freedom contributes to the pressure of hard spheres as $\alpha$=1 in this case. We can understand the contribution of the orientational fluctuations to the pressure using the additive approximation for the contact distance. For oblates HGO particles, the substitution of $\sigma(\theta)=a+2b\theta^2$ into Equation (14) yields

$$\lambda_0 = 4\pi \frac{e^{-Pa}}{P} \int\limits_0^{\pi/2} d\theta \sin\theta\, e^{-2Pb\theta^2} \qquad (25)$$

At high pressure, small-angle fluctuations dominate, permitting the approximation ($\sin\theta \approx \theta$) near $\theta$=0 and replacing the upper boundary of the integral with infinity. These approximations simplify Equation (25) to

$$\lambda_0 = \frac{\pi}{b} \frac{e^{-Pa}}{P^2} \qquad (26)$$

Substituting Equation (26) into $1/\rho = \partial(G/N)/\partial p$, yields the following equation of state: $P=2\rho/(1-\rho a)$. As $P_{||}=\rho/(1-\rho a)$ with $a = \sigma_{ee}$ for oblate particles, the pressure ratio becomes 2 in the vicinity of close packing density, i.e., $\alpha$=2. This shows that the close packing behavior of oblate HGO particles can be reproduced with the additive contact distance. Regarding the prolate particles, substituting $\sigma(\theta)=a+2b((\pi/2-\theta)^2$ into Equation (14), we can write that

$$\lambda_0 = 4\pi \frac{e^{-Pa}}{P} \int\limits_0^{\pi/2} d\theta \sin\theta e^{-2Pb(\pi/2-\theta)^2} \qquad (27)$$

Here, the definite integral can be evaluated in the limit of $P \rightarrow \infty$. It can be shown that

$$\lambda_0 = \frac{2\pi^{3/2}}{\sqrt{2b}} \frac{e^{-Pa}}{P^{3/2}} \qquad (28)$$

From this form, the corresponding equation of state can be written as $P=1.5\rho/(1-\rho a)$, where $a=\sigma_{ss}$. As the equation of state of orientationally frozen hard prolate particles is given by $P_{\parallel}=1.5\rho/(1-\rho a)$, we get that $\alpha=1.5$ for prolate HGO particles. This result is identical with the numerically obtained value of $\alpha$. Therefore, our analytical results for $\alpha$ confirm the validity of the additive approximation for the contact distance in describing the high-density behavior of HGO particles.

**Conclusion**

In this study, we have studied the orientational ordering and close-packing behavior of quasi-one-dimensional (q1D) hard Gaussian overlap (HGO) particles using the transfer operator method. Our study revealed the significance of particle shape anisotropy on the structural and thermodynamic properties. Oblate particles ($k<1$) are preferentially aligned with their short symmetry axes along the confinement direction ($z$-axis), giving rise to a perfect nematic ordering at high densities (pressures). In contrast, prolate particles ($k<1$) adopt orientational ordering perpendicularly to the confining $z$-axis, forming a planar alignment without in-plane orientational ordering even at the close-packing density. This means that prolate particles can form only a partially ordered structure. The difference between oblate and prolate shapes is clearly reflected in the nematic order parameter, which approaches unity for oblate particles but saturates at -1/2 for prolate ones, confirming the formation of perfect nematic order in one case and the partially ordered planar ordering in the other. Common in oblate and prolate shapes is that the orientational ordering happens in the polar angle ($\theta$) as oblate particles order into the preferred angle of $\theta_p=0$, while prolate ones into $\theta_p=\pi/2$. Regarding the azimuthal angle ($\varphi$) of the particle, it does not play a role in the nematic and planar ordering as $\theta_p=0$ selects only one special orientation, while $\theta_p=\pi/2$ is not accompanied by in-plane ordering in $\varphi$ angle. These results are due to the fact that the contact distance depends only on the relative azimuthal angle of two particles ($\varphi_{12}=\varphi_1-\varphi_2$). Considering the orientational fluctuations in $\theta$, the average of mean square orientational fluctuations vanishes with the same $\beta$ exponent for both shapes, i.e., $<(\theta_p-\theta)^2>\sim P^{\beta}$ where $\beta=-1$. Analytical treatments using the additive approximation for the HGO contact distance capture these trends accurately, particularly near the close-packing regime. Moreover, the orientational correlation length remains very short across the entire density range, indicating that strong ordering is a consequence of local, nearest-neighbor packing rather than long-range cooperative effects. This is consistent with the additive contact distance model, which results in $\xi=0$, i.e., the particles are uncorrelated in orientation. The close packing analysis of the pressure further reveals that orientational fluctuations have different contributions to the pressure for oblate and prolate shapes. For oblate particles, the contribution of orientational freedom equals that of translational freedom as $\alpha=2$, while for prolates, it is reduced to half as $\alpha=1.5$, and for spheres, orientational contributions vanish entirely as $\alpha$=1. These results can be understood based on the coupling between positional and orientational degrees of freedom. If two oblate particles are in contact and one of them tries to rotate by a general angle around the $x$ or the $y$ axes, which angle is denoted by $d\phi$, the rotating particle will collide and displace its neighbour by a distance $dz \sim (d\phi)^{1/2}$. Therefore the orientational fluctuations around the $x$ and $y$ axes each give $P^{-1/2}$ factor in $\lambda_0$, and ½ contribution to $\alpha$. However, the orientational fluctuations around the $z$ axes are not coupled to the positional fluctuations, i.e. the contact distance remains the same, while the particles rotate around $z$-axis. Therefore, these fluctuations have no additional contribution to the pressure and we obtain the above result for $\alpha$. Contrary, in the system of prolate particles neither the fluctuations around $z$ nor around the main axis of the particle, are coupled to the positional fluctuations, therefore they have no contribution to the pressure. However, the third orientational degree of freedom, namely the orientational fluctuations around the axis perpendicular to both $z$-axis and the main axis of the particle, gives a contribution to the pressure. As a result, the extra pressure due to the orientational fluctuation gives only 1/2 increment in $\alpha$. Finally, in the case of spheres the orientational and positional fluctuations are completely decoupled, therefore there is no extra increment and $\alpha$ remains 1. Taking into account the above results, it is not surprising that prolate shapes belong to the class of 2D hard bodies, where particles can perform rotation only in one angle. Therefore, the close packing rules of 2D hard bodies are also valid for 3D prolate particles, i.e., $\alpha+\beta$=1/2 and $\beta+\gamma$=-1. The oblate particles do not belong to this class, as perfect orientational ordering develops with increasing pressure. For oblate HGO particles, $\alpha+\beta$=1 and $\beta+\gamma$=-1 are obtained using both numerical and analytical calculations.

In summary, we have shown that the q1D system of 3D anisotropic hard bodies behaves very similarly to that of 2D anisotropic hard bodies, as $P/P\parallel$ indicates a structural change from a weakly ordered to a strongly ordered structure, and the close packing properties are independent of the aspect ratio. Our study showed that oblate systems do not belong to the class of 2D hard bodies, whereas prolate ones do. It remains an open question whether our close packing rules are valid for other hard bodies than HGO. To resolve this issue, studies with other prolate and oblate shapes such as cylinders, spherocylinders, and cut spheres, are needed.

## Acknowledgement

S.M. acknowledges financial support from the Alexander von Humboldt Foundation. P.G. and S.V. gratefully acknowledge the financial support of the National Research, Development, and Innovation Office - Grants No. 2023-1.2.4-TÉT-2023-00007 and No. TKP2021-NKTA-21. We thank Martin Oettel for his useful comments on this work.